\documentstyle[emulateapj,psfig]{article}
\def\gs{\mathrel{\raise0.35ex\hbox{$\scriptstyle >$}\kern-0.6em 
\lower0.40ex\hbox{{$\scriptstyle \sim$}}}}
\def\ls{\mathrel{\raise0.35ex\hbox{$\scriptstyle <$}\kern-0.6em 
\lower0.40ex\hbox{{$\scriptstyle \sim$}}}}
\submitted{Received 1998 May 26th; accepted: 1998 -- --}

\lefthead{Smail et al.}
\righthead{The Sub-millimeter Galaxy Population}

\begin{document}

\title{The Faint Sub-millimeter Galaxy Population: {\em Hubble Space
Telescope} Morphologies and Colors\footnotemark}

\author{Ian Smail,$\!$\altaffilmark{2,3} R.\,J.\ Ivison\altaffilmark{3,4},
A.\,W.\ Blain\altaffilmark{5} \& J.-P.\ Kneib\altaffilmark{6}}
\affil{\tiny 2) Department of Physics, University of Durham, South Road, 
Durham DH1 3LE, UK}
\affil{\tiny 4) Institute for Astronomy, Dept.\ of Physics \& Astronomy, 
       University of Edinburgh, Blackford Hill, Edinburgh EH9 3HJ, UK}
\affil{\tiny 5) Cavendish Laboratory, Madingley Road, Cambridge
CB3 OHE, UK}
\affil{\tiny 6) Observatoire Midi-Pyr\'en\'ees, 14 Avenue E.\ Belin,
F-31400 Toulouse, France}

\footnotetext{Based on observations with the NASA/ESA {\it Hubble
Space Telescope} obtained at the Space Telescope Science Institute,
which is operated by the Association of Universities for Research in
Astronomy Inc., under NASA contract NAS 5-26555.} 
\setcounter{footnote}{6}

\altaffiltext{3}{PPARC Advanced Fellow.}

\begin{abstract}
We present optical morphologies obtained from deep {\it Hubble Space
Telescope} ({\it HST}\,) and ground-based images for galaxies selected
from the first sub-millimeter (sub-mm) survey of the distant Universe.
Our sample comprises galaxies detected in deep 850-$\mu$m continuum
maps of seven massive clusters, obtained using SCUBA, the new bolometer
camera on the JCMT.  The survey covers a total area of 0.01\,degree$^2$
to 1$\sigma$ noise levels of about 2\,mJy\,beam$^{-1}$. We detect a
total of 25 sources at 850\,$\mu$m, of which 17 and 10 are brighter
than the respective 50\% and 80\% completeness limits.  Optical
counterparts are identified for 14 of the 16 sources in the $f_{50\%}$
sample and for 9 of the 10 sources in the $f_{80\%}$ sample that lie
within our optical fields.  The morphologies of those galaxies for
which we have {\it HST} imaging fall into three broad categories: faint
disturbed galaxies and interactions; faint galaxies too compact to
classify reliably; and dusty, star-forming galaxies at intermediate
redshifts.  The disturbed and interacting galaxies constitute the
largest class, which suggests that interactions remain an important
mechanism for triggering star formation and the formation of
ultraluminous galaxies in the distant Universe. The faint, compact
galaxies may represent a later evolutionary stage in these mergers, or
more centrally-concentrated starbursts. It is likely that some of these
will host active galactic nuclei.  Analysis of the colors of our sample
allow us to estimate a crude redshift distribution: $\gs 75$\% have
$z\ls 5.5$ whilst $\gs 50$\% lie at $z\ls 4.5$, suggesting that the
luminous sub-mm population is coeval with the more modestly
star-forming galaxies selected by UV/optical surveys of the distant
Universe.  This imposes important constraints on models of galaxy
formation and evolution.
\end{abstract}

\keywords{cosmology: observations --- cosmology: early universe ---
galaxies: evolution --- galaxies: formation --- galaxies: morphology}

\section{Introduction}

Recent technological advances in ground-based sub-mm technology
(Holland et al.\ 1998) are set to revolutionise our understanding of
the star-formation history of galaxies (Blain et al.\ 1998a, BSIK).
Deep maps of the sub-mm sky (Smail, Ivison \& Blain 1997, SIB) have now
resolved the population of galaxies responsible for the far-infrared
background (Puget et al.\ 1996; Fixsen et al. 1998; Schlegel, Finkbeiner
\& Davis 1998) and offer a precise determination of the star-formation
rate (SFR) in distant, dust-obscured systems (e.g.\ Dey et al.\ 1998).
This population is responsible for a substantial fraction of the total
SFR in the early Universe (BSIK) and holds crucial information about
how galaxies formed and evolved (Blain \& Longair 1993). 

Our sub-mm survey, the first results of which were reported by
SIB, exploits massive, foreground clusters to enhance the sensitivity
of the maps to distant star-forming galaxies (Blain 1997). The survey
has now been extended to cover seven clusters (Smail et al.\ 1998b, S98)
with sufficient sensitivity to detect a luminous infrared galaxy such
as Arp\,220 out to $z\sim 10$. Our sub-mm galaxy counts greatly
exceed those expected in a non-evolving model based on the local {\it
IRAS} 60-$\mu$m luminosity function (Saunders et al.\ 1990), or in a
model that can explain the optically-selected Lyman-dropout (Madau et
al.\ 1996) and Lyman-emission samples (Hu, Cowie \& McMahon 1998).

%
%
\begin{table*}
{\scriptsize
\begin{center}
\centerline{Table 1}
\vspace{0.1cm}
\centerline{Log of the SCUBA observations}
\vspace{0.3cm}
\begin{tabular}{lcccccccc}
\hline\hline
\noalign{\smallskip}
 {Cluster} & {R.A.} & {Dec.} & {$z$} &
  {Exposure time} & {$f_{850}(50\%)$} & {$f_{850}(80\%)$} & \multispan2{\hfil N ($f_{850}>f_{lim}$) \hfil } \cr
~ & {(J2000)} & {(J2000)} &  
~ & {(ks)} & {(mJy)} & {(mJy)} & {50\%} & {80\%} \cr
\hline
\noalign{\smallskip}
Cl\,0024$+$16 & 00~26~35.80 & $+$17~09~41.0 & 0.39 & 15.6 & 4.6 & 6.4 & 4 & 1 \cr
A370          & 02~39~53.00 & $-$01~35~06.0 & 0.37 & 33.8 & 5.7 & 7.2 & 3 & 3 \cr
MS\,0440$+$02 & 04~43~09.00 & $+$02~10~19.9 & 0.19 & 35.8 & 4.4 & 6.0 & 2 & 1 \cr
Cl\,0939$+$47 & 09~42~56.38 & $+$46~59~10.4 & 0.40 & 30.1 & 5.8 & 7.4 & 1 & 1 \cr  
A1835         & 14~01~02.20 & $+$02~52~43.0 & 0.25 & 23.0 & 5.0 & 6.6 & 3 & 2 \cr  
A2390         & 21~53~36.89 & $+$17~41~45.8 & 0.23 & 33.7 & 6.6 & 7.9 & 2 & 1 \cr
Cl\,2244$-$02 & 22~47~11.90 & $-$02~05~38.0 & 0.33 & 25.6 & 5.1 & 6.7 & 2 & 1 \cr
\noalign{\smallskip}
\noalign{\hrule}
\noalign{\smallskip}
\end{tabular}
\end{center}
}
\vspace*{-0.8cm}
\end{table*}

As demonstrated by BSIK, before we can understand the star-formation
history of the Universe, we need one remaining piece of the puzzle: the
redshift distribution, $N(z)$, of a faint sub-mm-selected galaxy sample.
The identification of the complete S98 catalog is underway (Ivison et al.\
1998a), this benefits from archival {\it HST} images for most of
the survey area as well as the availability of 10-$\mu$Jy\,beam$^{-1}$
radio maps for all the fields (Ivison et al.\ 1998b).  In the meantime,
useful limits on the sub-mm $N(z)$ can be determined from broad-band
imaging (Steidel et al.\ 1996).  Even a basic analysis along these lines
can challenge some currently viable models of galaxy evolution (BSIK).

In this paper we present limits on the redshift distribution of a
sample of sub-mm-selected galaxies derived from deep optical imaging.
We also take advantage of archival {\it HST} images to investigate the
optical morphologies for a sub-sample of our catalog on angular scales down
to 0.1\,arcsec ($\ls 1 h^{-1}_{50}$\,kpc at the expected redshifts).
This detailed morphological information can provide a valuable insight
into the physical mechanisms responsible for the high SFR of these
distant galaxies and thereby providing a fuller understanding of the
early stages of galaxy formation and evolution.

\section{Observations, Reduction and Analysis}

\subsection{Sub-millimeter Observations}

Our 850-$\mu$m maps were constructed using the long-wavelength array of
the Sub-millimeter Common-User Bolometer Array (SCUBA, Robson et al.\
1998) on the James Clerk Maxwell Telescope (JCMT)\footnote{The JCMT is
operated by the Observatories on behalf of the UK Particle Physics and
Astronomy Research Council, the Netherlands Organization for Scientific
Research and the Canadian National Research Council.}.  The final
on-source integration times and field centers are listed in Table~1.
Each field covers an area of 5.2\,arcmin$^2$, giving a total survey
area of 0.01\,degree$^2$.   The typical amplification for background
sources detected in our fields is expected to be $1.5^{+5.0}_{-0.5}$,
and so we have in effect surveyed an area of about 25\,arcmin$^2$ in
the source plane to an equivalent $1\sigma$ sensitivity of 1.3\,mJy. 

Source catalogs were constructed using the SExtractor package (Bertin
\& Arnouts 1996) and Monte Carlo incompleteness simulations undertaken
as described in SIB, for more details see Blain et al.\ (1998b) and
S98.  The derived sensitivity limits, 850-$\mu$m fluxes ($f_{850}$) for
50\% and 80\% completeness of $f_{850}(50\%)$ and $f_{850}(80\%)$, and
the number of detected sources in each field are listed in Table~1.  We
identify only two sources by applying the same selection criteria to
the negative fluctuations in our maps, both have
$f_{850}<f_{850}(50\%)$, indicating a false detection rate in our full
catalog of $\ls 10$\%. In the remainder of our analysis we will focus
on the two sub-samples from our full catalog of 25 sources:  10 sources
with $f_{850}>f_{850}(80\%)$ which we will denote $f_{80\%}$ and a
slightly more liberal sub-sample ($f_{50\%}$) defined by
$f_{850}>f_{850}(50\%)$ and containing 17 sources.  The $f_{80\%}$ and
$f_{50\%}$ limits are roughly equivalent to 4- and 3-$\sigma$ of the
sky noise.

\subsection{Archival Optical Observations}

In Table~2 we list the {\it HST} WFPC2 and the wider-field,
ground-based observations used in our analysis.  The WFPC2 observations
are mainly from programs with which we are associated; the remainder
were retrieved from the ST-ECF archive.  All were reduced as described
by Smail et al.\ (1997).  For one cluster, Cl\,0939$+$47, we have used
the deconvolved pre-refurbishment WF/PC-1 images of Dressler et
al.\ (1994) to determine the morphologies for SCUBA sources lying
outside the WFPC2 field of view.  The ground-based observations come
from a variety of sources, predominantly from programs on the 3.6-m
CFHT (Kneib et al.\ 1993) and the 5.1-m Hale (Dressler \& Gunn 1992;
Smail et al.\ 1998a).  The reduction and calibration of these images is
documented in those papers.  While lacking the spatial resolution and
depth of the WFPC2 images, these data have a wider
field of view and a more comprehensive wavelength coverage.
 
%
%
\begin{table*}
{\scriptsize
\begin{center}
\centerline{Table 2}
\vspace{0.1cm}
\centerline{Log of archival optical observations}
\vspace{0.3cm}
\begin{tabular}{lclcl}
\hline\hline
\noalign{\smallskip}
 {Cluster} &  {\it HST}/WFPC2  & Reference & Ground-based  & Reference \cr
~ & (Filter\,[Exposure time/ks]) &  &  \cr
\hline
\noalign{\smallskip}
Cl\,0024$+$16 &  F450W\,[23.4] F814W\,[13.2] & Smail et al.\ (1997)    & $I$       & Mellier (priv.\ comm.) \cr
A370          &  F336W\,[27.8] F675W\,[5.6]  & Archive                 & $UBV\!RI$ & Kneib et al.\ (1993) \cr
MS\,0440$+$02 &  F702W\,[18.4]               & Archive                 & ...       & ~~~~~~... \cr
Cl\,0939$+$47 &  F702W\,[21.0] F702W\,[22.0]$^a$  & Smail et al.\ (1997) & $gri$     & Dressler \& Gunn (1992) \cr  
A1835         &   ...                        & ~~~~~~...                     & $UBI$     & Smail et al.\ (1998) \cr  
A2390         &  F555W\,[8.4] F814W\,[10.5]  & Kneib et al.\ (1998)    & $UBI$     & Smail et al.\ (1998) \cr
Cl\,2244$-$02 &  F555W\,[8.4] F814W\,[10.5]  & Archive    & ...       &  ~~~~~~...\cr
\noalign{\smallskip}
\noalign{\hrule}
\noalign{\smallskip}
\multispan5{a) WF/PC-1 observations, reduced and deconvolved as described
in Dressler et al.\ (1994) \hfil} \cr
\end{tabular}
\end{center}
}
\vspace*{-0.8cm}
\end{table*}

Total magnitudes for all the sub-mm-selected galaxies covered by our
optical images were estimated using SExtractor's BEST\_MAG in the
reddest passband available for each cluster (either $R$ or $I$), using
{\it HST} frames if available. Galaxy colors were measured within
1-arcsec diameter apertures on the WFPC2 frames and 3-arcsec apertures
on the ground-based images, after matching the effective seeing on the
different frames.  If required, the photometry was converted to the
$I$-band assuming $(R-I)\sim 0.5\pm 0.2$, which is typical of both the
faint field population (Smail et al.\ 1995) and the available
photometry for our sub-mm counterparts. For Cl\,0939+47, the $(g-i)$
colors from Dressler \& Gunn (1992) were transformed to $(V-I)$ (Kent
1985).  No reddening corrections were applied.

\subsection{Optical Identifications}

The absolute positional accuracy of the SCUBA maps is better than
3\,arcsec, as estimated from repeated pointing checks performed during
the observations.  The positions of detected sources were determined by
fitting a circular gaussian to a 30-arcsec diameter region (two beam
widths) around the peak flux position on 4-arcsec-sampled maps.  The
random errors in the positions of individual objects depend on the
signal-to-noise ratio of the detections, but should be no worse than
4\,arcsec rms for the faintest sources discussed here.  The combined
positional uncertainty of our faintest sub-mm sources is therefore $\ls 6$\,arcsec (c.f.\ Fig.~1, Plate.~1).  Counterparts were
sought within this radius of the nominal sub-mm position. 

Using the deepest and/or reddest passband available for each cluster,
we identified all counterparts down to the magnitude limits of
our optical data for the 25 sub-mm sources from the full catalog.
These limits are typically $I\sim 23.5$ and $26.0$ for the ground-based
and {\it HST} exposures respectively.  Ten sources in the $f_{80\%}$
sample lie within our optical images; 7 of these are on the {\it HST}
frames and the remaining 3 on ground-based images. The equivalent
numbers for the 16 source in the $f_{50\%}$ sample are 11 and 5. In two
maps a {\em bright} sub-mm source has a clear counterpart, and
so their coordinates were corrected to align the sources exactly.

Due to the nature of sub-mm-selected galaxies, the optical counterparts
are likely to be faint, moreover the cluster fields analysed here are crowded.
Thus in some instances there is a non-zero likelihood of an unrelated
galaxy falling within the sub-mm error box.  As such our identifications
should be treated as preliminary until they have been confirmed from
deep radio maps, in particular it is possible that in a small number of
cases we have misidentified a sub-mm source due to the faintness of the
true counterpart combined with the presence of a nearby, brighter galaxy.
Nevertheless, as we show below the majority of the identified counterparts
are likely to be real and so the broad characteristics of luminous sub-mm
galaxies can be ascertained from our current catalog.

The likelihood of  a false identification for a given sub-mm source will
be a function of both the apparent magnitude of the identified counterpart
and its distance from the nominal position of the sub-mm source.  We have
used this information to estimate the probability of a real association
on an object-by-object basis.  Using a Monte Carlo simulation, based upon
the observed number counts of galaxies within each of our optical images,
we estimate the probability of a galaxy with the observed apparent
magnitude (or brighter) falling at random within a circle defined by
the optical and sub-mm positions.  The reliability of an identification
have been assigned to one of four classes based upon the probability of
a chance coincidence $P$: $P\leq 0.01$, definite; $P\leq 0.10$, likely;
$P\leq 0.25$, possible and $P\geq 0.25$, blank field.

\section{Results and Discussion}

The aim of this work is to obtain a first view of the optical
properties of the galaxies which contribute a substantial fraction of
the sub-mm emission in the Universe. To this end we retain the
$f_{50\%}$ sample although it is not statistically complete, instead we
use it to provide support for any trends seen in the smaller $f_{80\%}$
sample.  In the $f_{80\%}$ sample of 10 sources within the optical
fields, there are 4 counterparts with $P\leq 0.0$, 9 with $P\leq 0.25$
and 1 blank field. The rms offset between sub-mm sources and the 9
counterparts is $2.3\pm 0.5$\,arcsec, in reasonable agreement with the
nominal positional accuracy of the SCUBA maps.  The 16 sources with
$f_{850}>f_{850}(50\%)$ are associated with 6 definite identifications,
8 likely/possible identifications and 2 blank fields. 

In the following discussion we use all counterparts with $P\leq 0.25$, noting 
that none of our conclusions are substantially altered if we restrict our 
sample to those counterparts with $P\leq 0.10$.

Multicolor photometry is available for 9 of the 10 sources in the
$f_{80\%}$ sample (12/16 in the $f_{80\%}$). One of these 9 fields is
blank and in the remaining 8 we have 6 $B$ and 2 $V$-band detections.
For the more liberal $f_{50\%}$ sample we have 8 $B$-band and 4 $V$-band
detections from a total of 12 identified sources with multicolor
photometry. In all instances where we have both $B$ and $V$-band data,
the counterpart galaxies are detected in both bands.  The galaxies
plotted in Fig.~2 have a broad apparent-magnitude distribution  and
span the whole range of colors seen in the faint field population.  We
find no strong correlation between sub-mm flux and $I$-band
magnitude, as expected due to the wider selection function of a
sub-mm survey compared with that of optical surveys.

The detection of a galaxy in one of the bluer passbands can be used to
impose an upper limit to its redshift, because of the strong absorption
by the foreground Lyman-$\alpha$ forest at rest-frame wavelengths
shortward of 912\AA.  Thus the number of detections in the $B$- and
$V$-band images requires that at least 8/10 or $\gs 80$\% of the
galaxies in our $f_{80\%}$ sample have $z\ls 5.5$, the redshift at
which the Lyman-limit moves through the $V$-band. At least half of this
sample lies at $z\ls 4.5$ based on the $B$-band data alone. The
equivalent limits for the $f_{50\%}$ sample are $\gs 75$\% at $z\ls
5.5$ and $\gs 50$\% with $z\ls 4.5$.  Thus these luminous sub-mm
galaxies appear to span a similar redshift range to that covered by the
more modestly star-forming galaxies selected in UV/optical surveys of
the distant Universe (Madau et al.\ 1996).

%
%
\hbox{~}

\centerline{\psfig{file=f2.ps,angle=270,width=3.3in}}

\vskip 2mm

\noindent{
\scriptsize \addtolength{\baselineskip}{-3pt} 
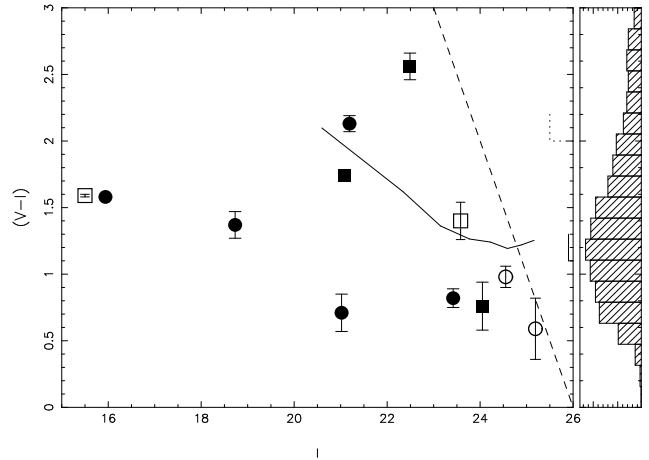
Fig.~2.\ The $(V-I)$--$I$ color-magnitude diagram for the 18
SCUBA sources for which we have multi-color information, the
approximate limits for the 2 non-detections are marked.  The filled
symbols and solid limits show those sources in the $f_{80\%}$ sample,
the open points and dotted limits are the remaining $f_{50\%}$
sources.  The square symbols represent $(B-I)$ measurements converted
to $(V-I)$ assuming $(B-V)=0.5$.   The apparent magnitudes of the
background sources have been corrected for a mean amplification of
1.5.  The dashed line shows the effective completeness limit in our
optical images. The solid line indicates the median $(V-I)$ color as a
function of $I$ magnitude for the faint field population, and
the histogram gives the distribution of $(V-I)$ colors for a sample of
field galaxies down to $I=25$ (Smail et al.\ 1995). 

\vskip 2mm
\addtolength{\baselineskip}{3pt}
}

The range of evolutionary models that can account for all the sub-mm and
far-infrared data currently available (BSIK) can already be constrained
by these crude limits. We can reject any model which predicts that more
than $\sim 20$\% of galaxies selected at an 850-$\mu$m flux limit of
3\,mJy lie beyond $z\sim 5.5$.  

Turning to the morphological composition of our catalog, we have used
high-quality {\it HST} images to provide detailed morphologies for
the 11 sub-mm-selected galaxies within these fields (Fig.~1).  For the
remaining 5 objects we use ground-based images, with typically sub-arcsec
resolution, to make cruder classifications.  Across the whole catalog we
find that the morphologies of the sub-mm galaxies can be split roughly
into 3 classes: 1) faint, disturbed galaxies, some of which are obviously
undergoing mergers and interactions; 2) faint, compact galaxies, too
small to classify reliably; and 3) bright, dusty star-forming galaxies. We
define a disturbed galaxy on the basis of the presence of one or more of
the following; multiple components, close companions with similar sizes
and colors, clearly distorted morphologies or tidal extensions.  Some of
these signatures are only visible on scales of $\ls 1$\,arcsec in the {\it
HST} images, and so there is an obvious bias against identifying such
systems in our ground-based data. The morphologies of the 9 identified
counterparts in the $f_{80\%}$ sample fall into the three classes in the
ratio 5(2):1(1):3(0), numbers in parentheses give the values from the
ground-based observations. The corresponding ratios for the 14 optical
counterparts in the $f_{850}>f_{850}(50\%)$ sample are 6(3):4(1):4(1).

The bright galaxies, shown in the bottom row of Fig.~1, include two
central galaxies from the clusters used in our survey.  These clusters
both contain strong cooling flows and their central galaxies show
signatures of strong star-formation activity and some non-thermal
emission.  The properties of these two galaxies are investigated in
more depth in a forthcoming paper. As confirmed cluster members these
two galaxies are removed from the analysis here. The remaining 2 bright
galaxies are both dusty Sab--Sbc spirals, lying at intermediate
redshifts,  within or in front of the clusters. The flux density
limit of our 850-$\mu$m survey means that we would expect to detect
this type of galaxy, with a SFR $\gs 20$--50$\,M_\odot$\,yr$^{-1}$, at
$z \ls 0.2$--0.5. We note that none of these bright galaxies was
included in the analysis of SIB. 

The optically faint sources are most interesting, as they typically lie
at high redshifts and have high SFRs, $\gs 100 M_\odot$\,yr$^{-1}$.
The number of these galaxies that are clearly disturbed/interacting
systems is roughly equal to the number that are either too faint or too
compact to be classified reliably.  This is a very robust result -- at
least 50\% of our faint, distant sub-mm-selected sources are identified
with distorted or interacting galaxies.  This interaction rate is
higher than the 30\% seen in the general field population at a similar
magnitude limit (van den Bergh et al.\ 1996).  The median separation of
the observed components in the interacting systems is about
$10h^{-1}_{50}$\,kpc. 50\% is thus a firm lower limit to the fraction
of interactions as some of the more compact sources, with typical
half-light radii of 0.2--0.4\,arcsec, or 2--$4 h^{-1}_{50}$\,kpc at
$z>1$, may be either unresolved or more evolved mergers. The exact
nature of the compact galaxies will require further investigation, in
particular we are obtaining high-resolution near-infrared images to
provide a more representative view of these galaxies.

Strong interactions and mergers are mechanisms responsible for the high
SFRs of ultraluminous infrared galaxies (ULIRG) in the local Universe
(Sanders \& Mirabel 1996).  The fraction of interacting ULIRGs found
locally is comparable to or greater than our lower limit, but the mean
separation of the components is close to the resolution limit of even
our WFPC2 images. Hence we suggest that the rate of interaction found
in our distant, sub-mm-selected galaxies is probably comparable to that
seen in local ULIRGs. On the basis of our morphological survey we can
state that one mechanism, dynamical interaction, is responsible for
triggering at least half of the sub-mm flux density, and by
implication for at least half of the star-formation activity in the
most distant strongly star-forming galaxies in our sample. These
observations provide the clearest support for the pivotal role of
mergers and interactions in the early evolution of luminous
star-forming galaxies.

\section{Conclusions}

\noindent{$\bullet$} We present optical identifications of galaxies
selected in a deep sub-mm survey.  Roughly 70\% of the sources selected
at 850\,$\mu$m, for which optical imaging data is available, have
definite/likely counterparts.  A further 20\% of the sub-mm sources
have probable optical counterparts, only 20\% of the sub-mm sources
have no identifiable counterpart to a typical apparent magnitude of
$I\sim 25$--26.

\noindent{$\bullet$} 75\% of the optical counterparts for which we have
$B$ or $V$-band images are detected in these passbands, suggesting
that at least three-quarters of the 850-$\mu$m-selected galaxies
brighter than 3\,mJy lie at $z\ls 5.5$.  

\noindent{$\bullet$} The morphologies of the fainter (and hence
probably more distant) optical counterparts fall roughly equally into
two classes: disturbed/interacting and compact objects.  The
prevalence of interactions in this survey underlines the central role
of this mechanism in triggering star formation in the most luminous
galaxies at high redshift.

\section*{Acknowledgements}

We thank the commissioning team of SCUBA and Ian Robson for his continuing
support. We also thank Yannick Mellier for allowing us to use his exquisite
CFHT images of Cl\,0024+16, and Wayne Holland, Tim Jenness, Malcolm
Longair and Michael Rowan-Robinson for useful conversations and help.
IRS acknowledges support from the Royal Society and the Australian
Research Council while a Visiting Fellow at UNSW, Sydney.

%
%
\hbox{~}\bigskip
\centerline{\psfig{file=f1.ps,width=6.0in}}
\smallskip

\smallskip
\noindent{\scriptsize
\addtolength{\baselineskip}{-3pt} 
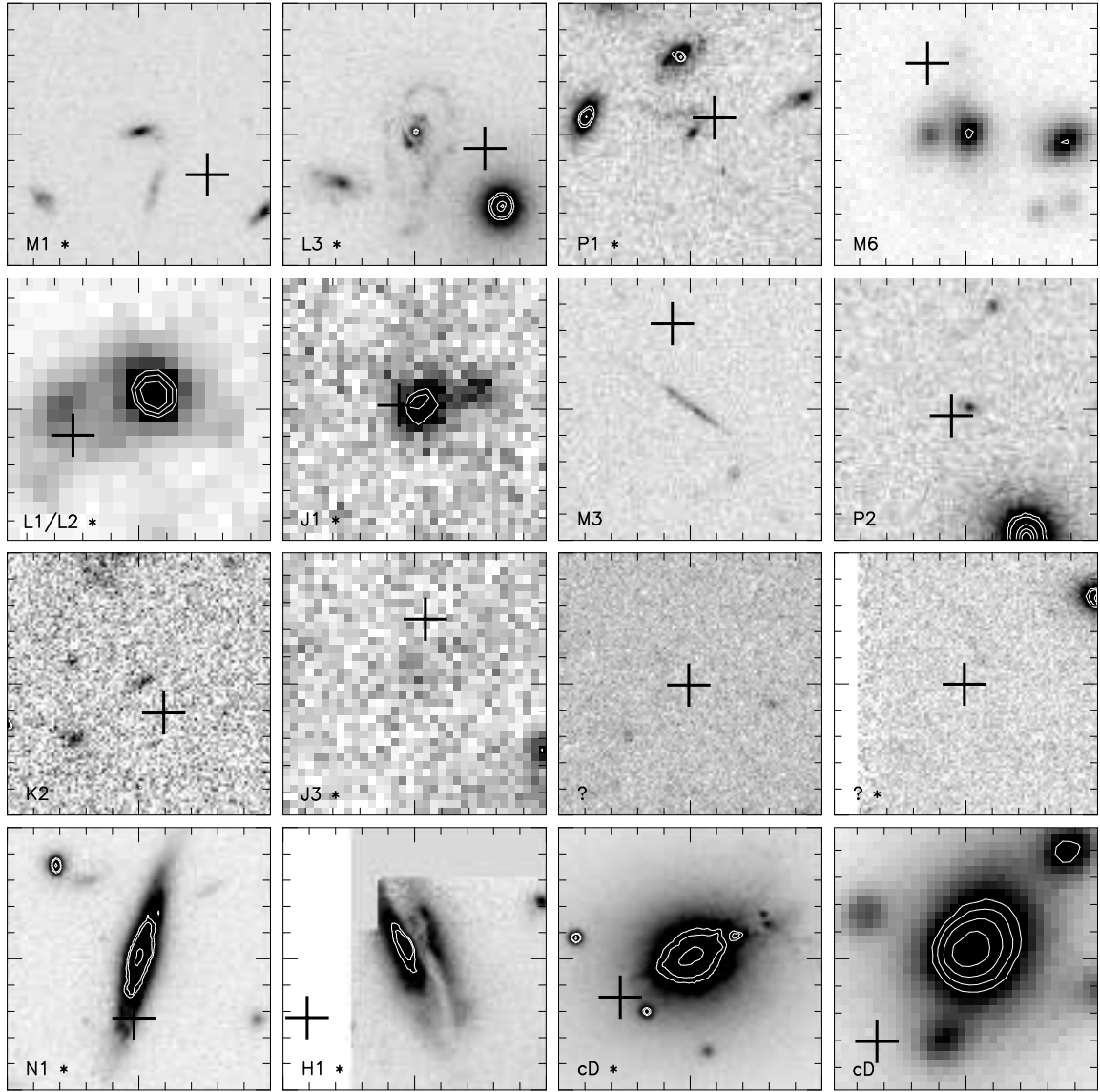
Fig.~1. $10\times 10$\,arcsec images of the 16 sub-mm sources in our
$f_{50\%}$ sample which fall within our optical imaging of the seven
clusters.  These are ordered from the upper-left on the basis of their
morphologies: 6 disturbed/interacting, 4 compact/featureless (including
a strongly-distorted arclet), 2 blank fields and then 4 bright, dusty
galaxies. A `?' next to the sub-mm source name indicates a blank field,
a `$\ast$' marks those sub-mm sources for which $f_{850}>f_{850}(80\%)$.
The images are centred on the most likely optical candidate. The
centroids of the sub-mm sources are indicated by crosses.  Note that
these images span a range in exposure times and resolutions (Table~2).
The panels correspond to $\gs 80 h^{-1}_{50}$\,kpc at $z>1$.  

\addtolength{\baselineskip}{3pt}
}

\end{document}